\documentclass[aps,prl,twocolumn,superscriptaddress,groupedaddress]{revtex4}  
\usepackage{graphicx}  
\usepackage{dcolumn}   
\usepackage{bm}        
\usepackage{amssymb}   
\usepackage{color}
\usepackage{amsmath}
\newcommand{\be}{\begin{equation}}
\newcommand{\bea}{\begin{eqnarray}}
\newcommand{\eea}{\end{eqnarray}}
\newcommand{\ba}{\begin{align}}
\newcommand{\ea}{\end{align}}
\newcommand{\ee}{\end{equation}}

\begin{document}

\widetext
\begin{flushright}
MPP-2019-119\\
IPM/P-2019/021\\
\end{flushright}

\title{Universal Scaling in Fast Quenches Near Lifshitz-Like Fixed Points}

\author{M. Reza Mohammadi Mozaffar} \email{mmohammadi@guilan.ac.ir}

\affiliation{Department of Physics, University of Guilan, P.O. Box 41335-1914, Rasht, Iran}
\affiliation{School of Physics, Institute for Research in Fundamental Sciences (IPM), P.O.Box 19395-5531, Tehran, Iran}

\author{Ali Mollabashi} \email{alim@mpp.mpg.de}

\affiliation{Max-Planck-Institut for Physics, Werner-Heisenberg-Institut 80805 Munich, Germany}

\date{\today}

\begin{abstract}
We study critical dynamics through time evolution of quantum field theories driven to a Lifshitz-like fixed point, with $z>1$, under relevant deformations. The deformations we consider are \textit{fast smooth} quantum quenches, namely when the quench scale $\delta t^{-z}$ is large compared to the deformation scale. We show that in holographic models the response of the system merely depends on the scaling dimension of the quenched  operator as $\delta\lambda\cdot\delta t^{d-2\Delta+z-1}$, where $\delta\lambda$ is the deformation amplitude. This scaling behavior is enhanced logarithmically in certain cases. We also study free Lifshitz scalar theory deformed by mass operator and show that the universal scaling of the response completely matches with holographic analysis. We argue that this scaling behavior is universal for any relevant deformation around Lifshitz-like UV fixed points.  
\end{abstract}

\maketitle
\section{Introduction}
Quantum quenches are powerful theoretical / experimental probes for respond of closed systems to time dependent couplings in the Hamiltonian. There are several motivations for studying such systems among which understanding \textit{critical dynamics} is a very interesting one. A quenched system driving to / crossing a critical point is an appropriate tool to study critical dynamics (see e.g. \cite{uQPT}).

Quantum quenches had shed light on distinct problems including relaxation process in integrable models \cite{Relaxation} and also is widely studied to understand thermalization process in strongly coupled theories \cite{thermalization}, thanks to gauge/gravity duality. In the \textit{slow} quench regime, several physical quantities obey Kibble-Zurek scaling \cite{KZ} near the critical point. Holographic quenches has also uncovered a new universality class in \textit{fast} quench regime  \cite{Buchel:2012gw, Buchel:2013gba} which was later showed to be a general feature of theories with a UV conformal fixed point \cite{fastFT}.  

A quantum critical point is characterised by two critical exponents. These two exponents correspond to divergence of the correlation length $\xi$, and vanishing of the energy scale of fluctuations $\delta$, during a second order phase transition. It turns out that $\delta\sim\xi^{-z}$ where $z$ is defined as the \textit{dynamical critical exponent}. Quantum critical points often have $z\neq 1$ (see e.g. \cite{Hertz:1976zz, QPT}).

In this letter we study fast smooth quenches for generic dynamical critical exponent ($z>1$). Smooth quenches are defined as $\delta t^{-1} \ll \Lambda$ where $\delta t$ is the duration of the quench and $\Lambda$ is the UV cut-off scale \footnote{In contrast instantaneous quenches are defined as $\delta t^{-1} \gtrsim \Lambda$ \cite{Relaxation}.}. A smooth quench is called \textit{fast} when the scale of the quench is small compared to any other scale, e.g. the initial gap, in the theory.

We consider the vacuum state of a theory with a Lifshitz-like fixed point in very distinct regimes, namely in (\textit{i}) a special class of strongly coupled theories with large number of degrees of freedom and in (\textit{ii}) free solvable Lifshitz scalar theories. We show that under relevant scalar deformations, these theories obey exactly the same universal scaling behaviors near the fixed point \footnote{Recently fast and slow quenches have been studied in non-relativistic fermionic models \cite{Das:2019qaj}}.

\section{Holographic Quenches}
We consider a strongly coupled theory admitting a Lifshitz-like UV fixed point. A reasonable choice as the gravity dual for such a theory is introduced in \cite{Taylor:2008tg}. The deformed theory with a relevant scalar operator with a time-dependent coupling $\lambda(t)$ is given by
\begin{align}\label{eq:HoloAction}
\begin{split}
S&=\frac{-1}{16\pi G_N}\int d^{d+1}x\sqrt{-g}
\Big(\mathcal{R}+\Lambda-\frac{1}{2}(\partial \chi)^2
\\&\hspace{9mm}
-\frac{1}{4}e^{\lambda \chi}F^2-\frac{1}{2}(\partial \phi)^2-\frac{1}{2}m^2\phi^2-V(\phi)\Big),
\end{split}
\end{align}
where the scalar $\chi$ and two-form $F$ are responsible for the dynamical exponent parameter in the solution space and $V(\phi)$ contains higher (than quadratic) powers of $\phi$. The scalar field $\phi$, with $m^2=\Delta(\Delta-d_z)$, is dual to the deformation operator on the boundary side with scaling dimension $\Delta$ and $d_z:=d+z-1$. Relevant deformations on the bulk side are equivalent to $m_{\text{BF}}^2<m^2<0$ where $m^2_{\text{BF}}=-d_z^2/4$ \cite{Andrade:2012xy}.

We are interested in the solutions where the geometry dual to the vacuum state of a theory with a Lifshitz-like fixed point is given by
\begin{equation}\label{eq:metric}
ds^2=-\frac{f(t,r)}{r^{2z-2}}dt^2+\frac{dr^2}{r^4f(t,r)}+g(t,r)^2d\vec{x}^2,
\end{equation}
where in the undeformed case we have $f(t,r)=r^{-2}$ and $g(t,r)=r^{-1}$ and we set the bulk radius scale to unity. The energy condition of the gravity theory restricts these solutions to have $z\geq 1$. It is worth to note that since our discussion is mainly at the level of geometry (i.e. dual state), it truly does not matter whether the underlying theory is given by \eqref{eq:HoloAction} or other theories admitting such a solution provided a very generic condition which we discuss bellow is satisfied.

In general dealing with a time dependent coupling even in the vacuum state is not an easy task.
Here the idea is very similar to what has been addressed in \cite{Buchel:2013gba} for quenching a deformed conformal field theory. We consider a time dependent source which is turned on at $t=0$ for a short duration $\delta t$ and is held fixed afterwards. The crucial point in the analysis is that on the bulk geometry the response starts to propagate after $t=0$ in a region restricted by $t=r^z/z$ and the asymptotic boundary $r\to0$. As long as $\delta t$ is small compared to other scales in the system, the interesting physics is happening in this region which is safe from backreaction. In other words, smallness of the quench rate with respect to other scales results in the non-trivial physics to happen before the geometry finds enough time to backreact on the response of the system to the deformation.

To put this argument in a more precise way we first define the following dimensionless hated coordinates/fields as $r=\delta t \,\hat{r}$, 
$t=\delta t^z\, \hat{t}$, $f=\hat{f}\,\delta t^{-2}$, $g=\hat{g}\,\delta t^{-1}$ and $\phi=\delta t^{d_z-\Delta} \, \hat{\phi}$ \footnote{Note that in this convention $t$ and $\delta t$ have different scaling dimensions.}. One can easily check that in the $\delta t\to 0$ limit, Einstein equations decouple from the scalar and the scalar equation turns out to be nothing but a scalar on pure Lifshitz geometry, namely  
\begin{align}\label{eq:scalarEq}
\begin{split}
&\hat{r}^2 \partial^2_{\hat{r}}\hat{\phi}-(d_z-1)\hat{r} \partial_{\hat{r}}\hat{\phi}-\hat{r}^{2z} \partial^2_{\hat{t}}\hat{\phi}+\Delta(d_z-\Delta)\hat{\phi}=0.
\end{split}
\end{align}
In the following analysis we restrict to cases where $2\Delta\neq d_z+2nz$ where $n$ is a positive integer. We discuss about these particular cases later on in this section. 

The scalar field near the boundary is given by
\begin{align}
\begin{split}
\phi(\hat{t},\hat{r})&=\delta t^{d_z-\Delta}\hat{r}^{d_z-\Delta}\left[p_s(\hat{t})+\mathcal{O}(\hat{r}^{2z})\right]
\\&\hspace{25mm}
+\delta t^{\Delta}\hat{r}^{\Delta}\left[p_r(\hat{t})+\mathcal{O}(\hat{r}^{2z})\right],
\end{split}
\end{align}
in terms of the source $p_s$ and the response $p_r$ functions. The corresponding coefficients at higher orders which we have not mentioned explicitly depend on $\Delta$, $d$ and $z$.

We consider a simple source function for $0<t<\delta t$ as
\be 
p_s(\hat{t})=\delta p\,{\hat{t}}^\kappa,
\ee
where $\kappa>0$ and the function is hold fixed with $p_s(\hat{t})=\delta p$ for $t>\delta t$. Since the dynamics of the response is restricted to the region between the Lifshitz boundary $r\to0$ and $t=r^z/z$, smoothness condition implies that $\phi(t=r^z/z,r)=0$. This leads to
\be\label{eq:HoloRes}
p_r(\hat{t})=a_\kappa\cdot\delta p\cdot\delta t^{d_z-2\Delta}\cdot\hat{t}^{\frac{d_z-2\Delta}{z}+\kappa},
\ee
where $a_\kappa$ is found from the solution of the scalar field on the entire region as
$$a_\kappa=-\frac{(2z)^{\frac{d_z-2\Delta}{z}}\Gamma(\kappa+1)\Gamma\left(\frac{d_z-2\Delta}{2z}+1\right)}{\Gamma\left(\frac{d_z-2\Delta}{z}+\kappa+1\right)\Gamma\left(\frac{2\Delta-d_z}{2z}-1\right)}.$$
From the above expression we find that the analysis is valid for $\kappa>-(d_z-2\Delta)/z$. This implies that for quenches which do not satisfy this constraint we are not allowed to ignore the backreaction of the geometry. Also it is notable that the above analysis is trivially generalized for more complicated source functions owing to the linearity of Eq.\eqref{eq:scalarEq}. 

In specific cases where $2\Delta=d_z+2nz$, holographic renormalization of scalar operator indicates that there is an extra logarithmic term in the expansion of the response function \cite{Andrade:2012xy}. In these cases the leading scaling behavior of the response is enhanced logarithmically as $p_r(\hat{t})\sim \delta t^{d_z-2\Delta} \log\delta t$.

For homogeneous quenches the diffeomorphism Ward identity, reads as $\partial_t\mathcal{E}+\langle\mathcal{O}\rangle\partial_t\lambda=0$ where $\mathcal{E}$ is the energy density of the boundary theory \cite{Buchel:2012gw}. The equation is the same as the constraint, namely the $\{r,t\}$ component, Einstein equation. From an integrated version of this identity
\be\label{eq:ward}
\Delta\mathcal{E}_\text{ren}\propto\int\,dt\,p_r\,\partial_t p_s
\ee
we find that the energy density obeys the same scaling behavior as the response function. This provides us a non-trivial check comparing our holographic and field theory result in the next section. 

It is worth to note that the above analysis is quite general in this sense that it is valid for any kind of non-linear quench with an arbitrary amplitude. Although we have considered a quench on the vacuum state, due to the region which the non-trivial dynamics of the response takes place, the analysis applies in the same manner to any initial state with such a UV structure. The analysis is applicable to any bulk theory which in the $\delta t\to0$ limit the Einstein equations decouple from the field dual to the deformation operator.

Our last comment in this section is regarding to non-uniqueness of the choice of boundary condition in gauge/gravity duality \cite{Klebanov:1999tb}. 
The so-called `alternative quantization' scenario is allowed for operators with $m^2_{\text{BF}}<m^2<m^2_{\text{BF}}+z^2$ \cite{Andrade:2012xy, 1212}. There is an extra instability which further constraints the scaling dimension by $\Delta\leq -1+d_z/2$ \cite{Andrade:2012xy}. In this restricted regime, the scaling of the response function turns out to be the inverse of that of standard quantization in Eq. \eqref{eq:HoloRes} and the logarithmic enhancement happens for $2\Delta=d_z+2n$ where $n$ is a non-negative integer.

It is hard to say how this universal behavior, namely that the scaling of the response function being independent of the quench details, depends on holographic nature of states or not. In the next section we study certain relevant deformations in free scalar theories with Lifshitz-like fixed points. We show that free field theory results matches with what we have found holographically, indicating that the universal behavior may be the case in a much wider sense. 

\section{Free Theory Quenches}
We now focus on the opposite limit, where the theory is non-interacting. We consider a mass quench for a free scalar theory in $d$-dimensions admitting a Lifshitz-like fixed point in the massless limit,
\be\label{eq:FTaction}
S=\frac{1}{2}\int dt d\vec{x} \left[\dot{\phi}^2-\sum_{i=1}^{d-1}(\partial_i^z \phi)^2-m^{2z}(t) \phi^2\right].
\ee
The quench operator we consider in this case is the mass operator $\phi^2$ with $\Delta=d-z-1$ and the time dependent coupling is $m^{2z}(t)$. For a given $z$ the mass operator is relevant for $d>z+1$. We are interested in a mass profile starting at some positive $m_0$ in $t<0$ and smoothly land on a Lifshitz invariant theory later on. The smoothness rate is again fast compared to initial gap $m_0^z$, and the UV cut-off scales. To deal with a solvable dynamic, we choose the following profile
\be\label{eq:FTmass}
m^{2z}(t)=\frac{m^{2z}_0}{2}\left(1-\tanh\frac{t}{\delta t^z}\right)
\ee
bringing the theory from a gapped phase to a gapless Lifshitz invariant one. Considering the following expansion for $\phi$
\be
\phi(x,t)=\int d^{d-1}k\left(a_k u_k + a^\dagger_k u^*_k\right),  
\ee
where $[a_k,a^\dagger_{k'}]=\delta^{d-1}(k-k')$, one finds that the in-going modes are given by
\begin{align}
\begin{split}
\hspace{-2mm}
u_k&=\frac{1}{\sqrt{2\omega_{\text{in}}}}e^{i(k.x-\omega_+t)}\left(2\cosh\frac{t}{\delta t^z}\right)^{-i \omega_-\delta t^z}\times
\\&
_2F_1\left(1+i \omega_-\delta t^z,i \omega_-\delta t^z,1-i \omega_{\text{in}}\delta t^z;1-\frac{m^{2z}(t)}{m_0^{2z}}\right),
\end{split}
\end{align}
where $\omega_{\text{in}}=\sqrt{k^{2z}+m_0^{2z}}$ and $\omega_{\pm}=(|k|^z\pm\omega_{\text{in}})/2$. We are interested in the (renormalized) expectation value of the mass operator given by
\be\label{eq:phi2ren}
\hspace{-3mm}
\langle\phi^2\rangle_\text{ren}=\sigma_d\int\,dk\left(\frac{k^{d-2}}{\omega_\text{in}}\left|_2F_1\right|^2-f^{(d)}_\text{ct}(k,z,m(t))\right),
\ee
where $\sigma_d^{-1}=2(2\pi)^d/\Omega_{d-2}$. The counter terms contribution is non-vanishing for $d\geq z+2$ where there exists divergent terms in the large momentum expansion of the integrand with $k^{d-z-2}$ being  the most divergent term. There are at least two ways to get rid of these divergent contributions. One way is to renormalize by subtracting the result from a slow quench, known as the adiabatic expansion \cite{KZ}. The reason why the adiabatic expansion is supposed to work is that the UV modes do not care about the details of the quench so they are supposed to have the same divergent structure in fast and slow quenches. For larger values of the dynamical exponent, these UV modes happen to become careless to the quench details relatively at smaller momenta values due to the $z$ power enhancement in the dispersion relation. The second alternative to get rid of these divergent contributions is to directly regularize the result by subtracting divergent terms appearing in the large momentum expansion of the integrand. These two methods give the same result in our case. A closed form for $f^{(d)}_\text{ct}$ which depends on $m^{2z}(t)$ and its time derivatives is found but is not much instructive to be presented here. It is also worth to note that for $d<3z+1$ there is a leading constant term which we absorb it in our choice of renormalization scheme. 

We analyze the fast quench regime with the mass operator first analytically at first non-trivial order of the $\delta t$-dependence of \eqref{eq:FTaction}. We find that the scaling dimensions found perturbatively perfectly agree with our fits to numerical analysis.

Lets first consider dimensionless hated parameters for the initial mass $\hat{m}_0=m_0\,\delta t$ and the momentum variable $\hat{k}=k\,\delta t$. In order to explore the fast quench we need to expand the hypergeometric function for a fixed $\hat{k}$ around $\hat{m}_0=0$. One can easily check that the leading order expansion is $\hat{m}_0^{2z}$. 

It happens for $d=z+1+2nz$ where $n$ is a non-negative integer that $\hat{k}^{-1}$ appears in this expansion. In these cases we need to add a IR cut-off $\mu$ to our analysis. In our numerical analysis we always set $\mu=1$. 
\begin{figure}
\begin{center}
\includegraphics[scale=0.3]{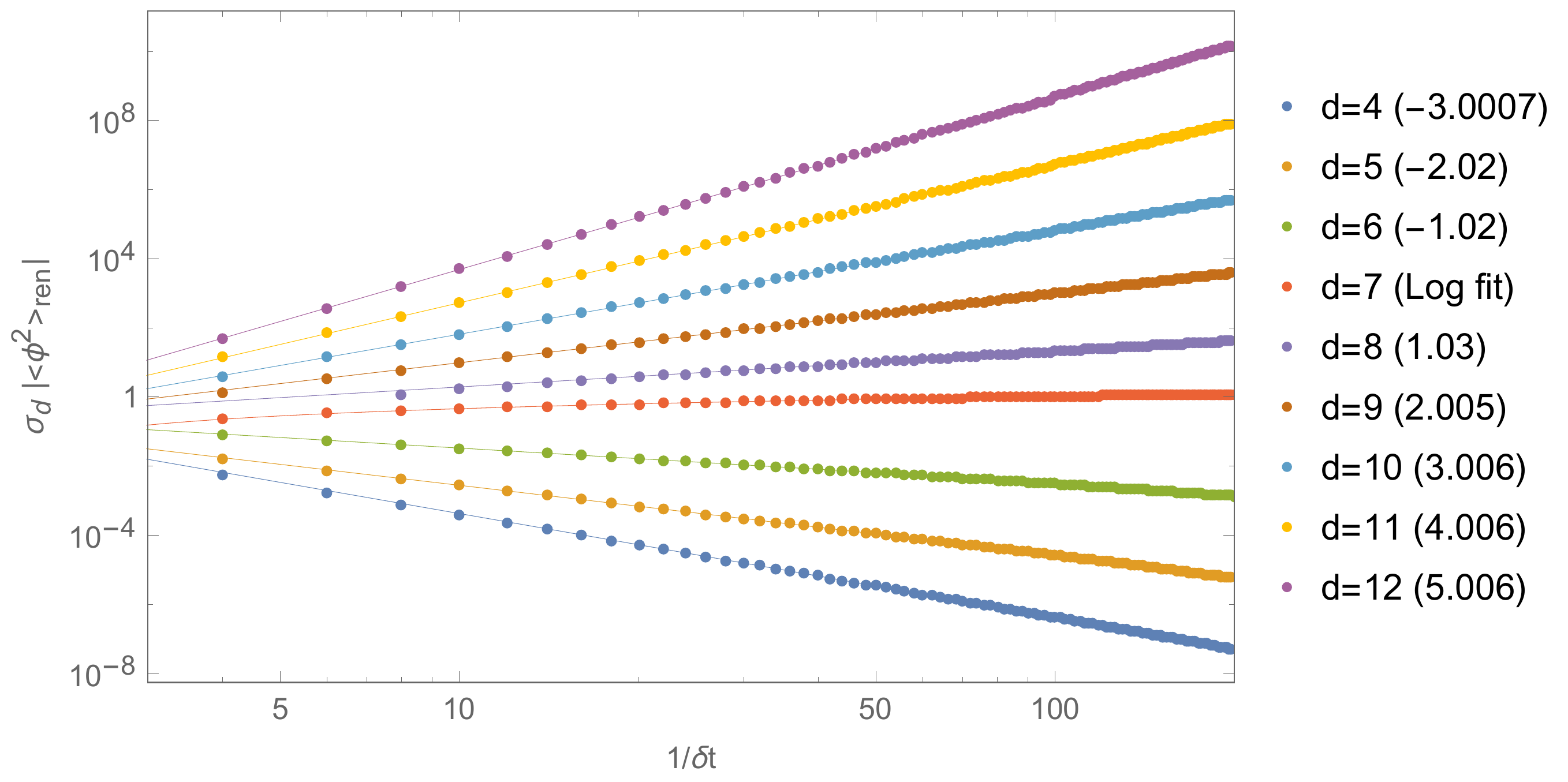}
\end{center}
\caption{Numerical data for the renormalized mass operator for $z=2$ (we set $t=0$). The solid lines correspond to power law fit functions with given exponents in the plot legend for $d\leq 4$ where $\phi^2$ is relevant for $z=2$.}
\label{fig:scaling}
\end{figure}

The analysis of \eqref{eq:phi2ren} shows that the generic form is given by
\be\label{eq:FTgenRes}
\langle\phi^2\rangle_\text{ren}=c_d\cdot m_0^{2z}\cdot\delta t^{3z+1-d}+\mathcal{O}\left(\delta t^{5z+1-d}\right),
\ee
where $c_d$ is a dimensionless function of the dynamical exponent and the mass profile. The exact form of $c_d$ does not really matter in our analysis (see \cite{MMinProgress} for a more detailed discussion), though for specific case $d=2z(2+n)+1$, where $n$ is a non-negative integer, the leading term in \eqref{eq:FTgenRes} takes the following nice form 
\be
\frac{(-1)^{n}}{2^{3+2n}}\frac{\pi}{z}\partial_t^{2n+1}m^{2z}(t).
\ee
This scaling result Eq.\eqref{eq:FTgenRes} matches with the holographic result Eq.\eqref{eq:HoloRes}. 

\begin{figure}
\begin{center}
\includegraphics[scale=0.35]{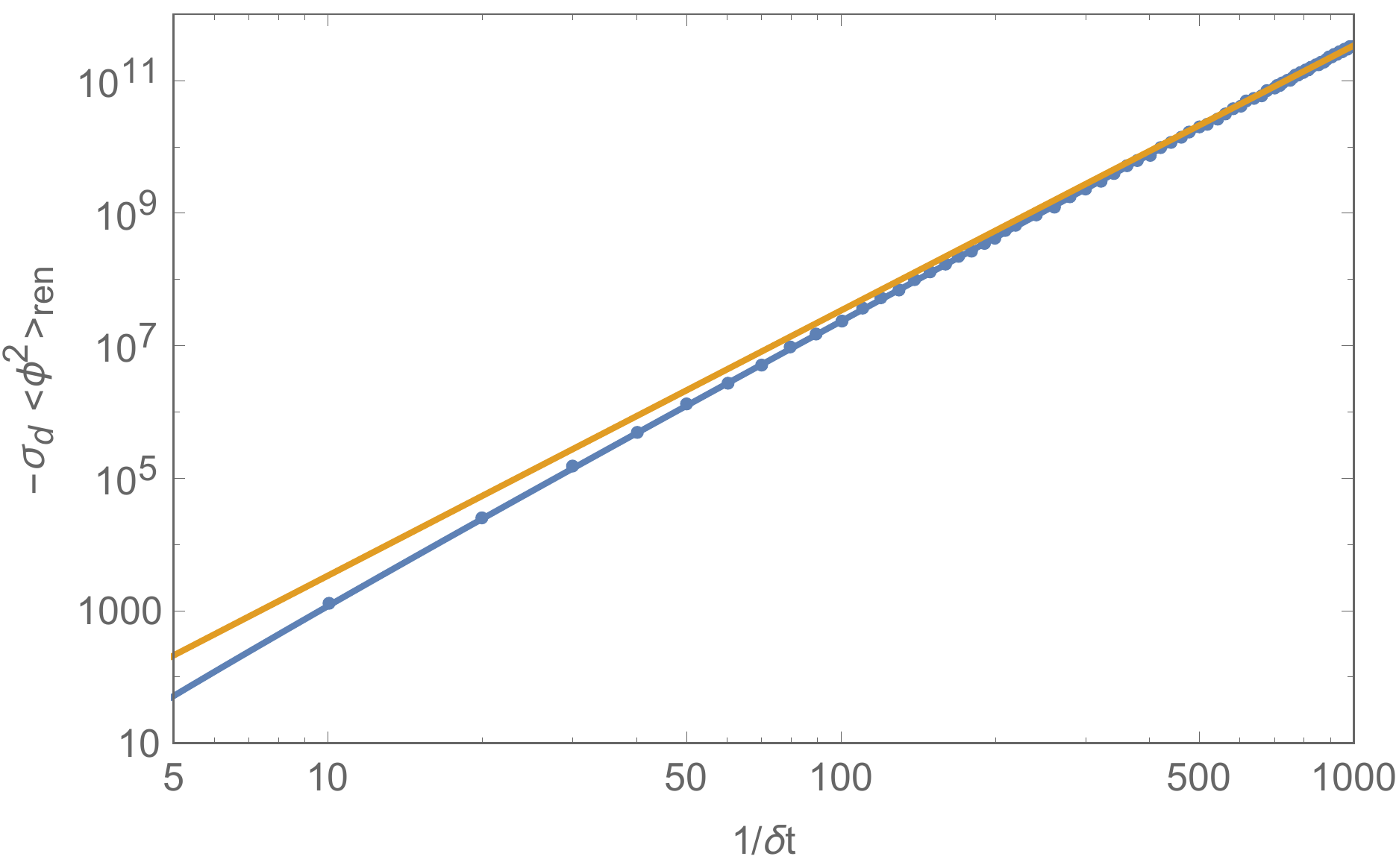}
\end{center}
\caption{Logarithmic enhancement of scaling for $d=11$, $z=2$, and $t=\delta t^z/2$. The blue curve is fitted with $\propto\delta t^{\alpha} \log\delta t$ (where the best fit gives $\alpha=3.993$) and the orange line is the best fit with $\propto\delta t^{3d+1-d}$.}
\label{fig:logEnhanc}
\end{figure}
In Fig.\ref{fig:scaling} we present numerics for Eq.\eqref{eq:phi2ren} and the corresponding power-law fits. The exponents we find numerically perfectly matches with Eq.\eqref{eq:FTgenRes} and thus with holographic results. In addition, the holographic analysis predicts logarithmic leading behavior for special case with $2\Delta=d_z+2nz$. For the case of relevant mass operator the condition is nothing but $d=3z+1+2nz$ for non-negative integer $n$. For the case of $n=0$ this is in agreement with Eq.\eqref{eq:FTgenRes} when the exponent of $\delta t$ vanishes. In our numerics for the case of $z=2$, as shown in Fig.\ref{fig:scaling}, we found the best fit to be logarithmic for $d=7$. We find that the coincidence between field theory and holographic results is also correct for $n>0$. For instance in the case of $z=2$ the next value of $d$ turns out to be $d=11$ while the data in Fig.\ref{fig:scaling} does not show such a behavior! The reason is due to our specific choice for the mass profile Eq.\eqref{eq:FTmass} where this enhancement disappears at $t=0$ but shows up for other values of $t$. In Fig.\ref{fig:logEnhanc} we have plotted numerical data for such a case where the logarithmically enhanced fit perfectly matches the data points. We have found the same sensitivity of numerics for larger values of $n$ and $z$ and always found perfect agreement with the holographic prediction. 

\begin{figure}
\begin{center}
\includegraphics[scale=0.4]{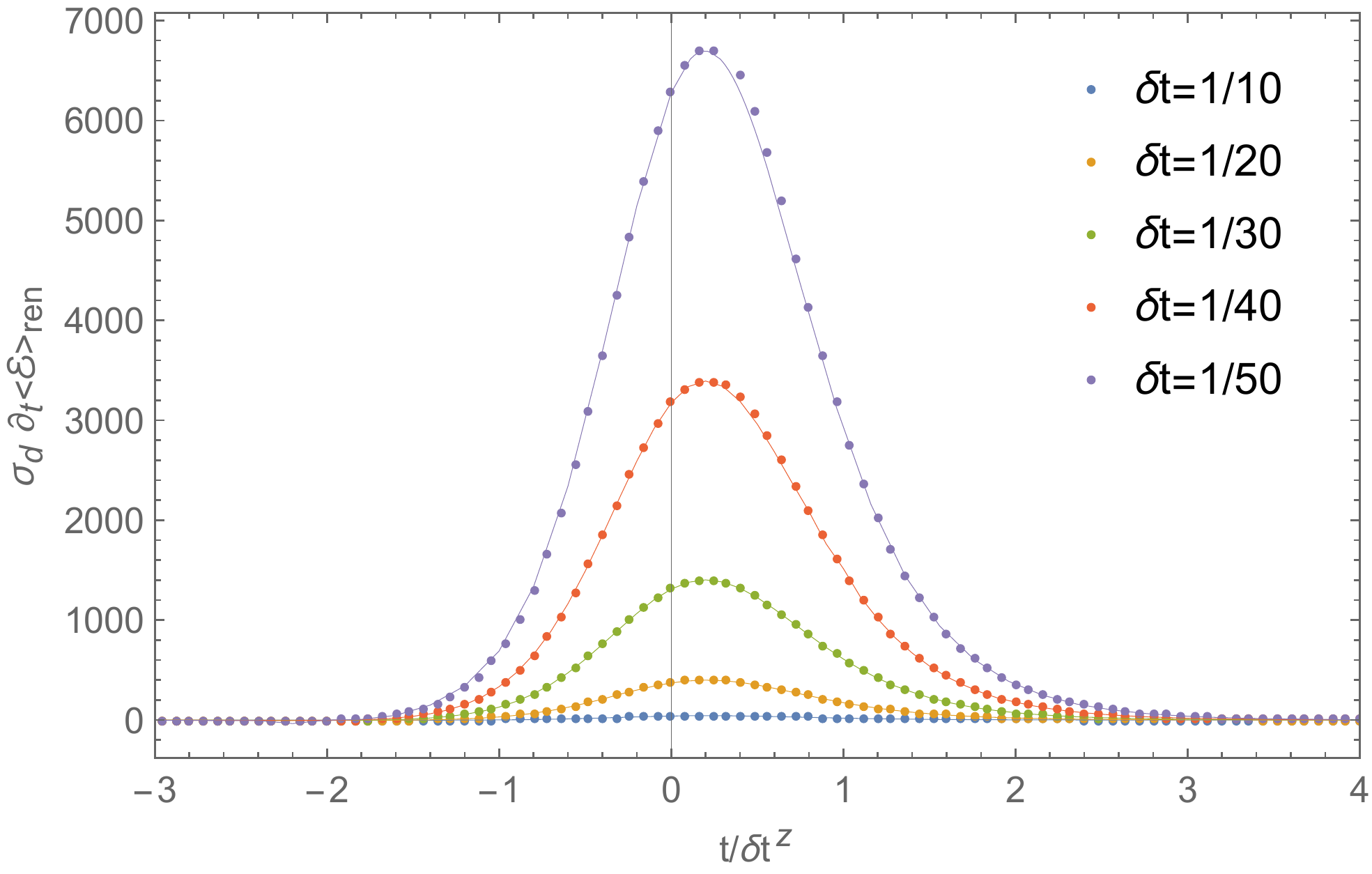}
\end{center}
\caption{Solid line is $\partial_t \mathcal{E}_\text{ren}$ and the data points are $-\frac{1}{2}\partial_t m^{2z}(t)\langle\phi^2\rangle_\text{ren}$ presented for different values of $\delta t$. We set $d=8$ and $z=2$.}
\label{fig:ward}
\end{figure}

In our holographic analysis using Ward identity (constraint Einstein equation) we showed that the energy density and the response function scale identically. As our final check for our results we check this identity, namely the renormalized version $\partial_t\mathcal{E}_\text{ren}=\frac{1}{2}\partial_t m^{2z}(t)\langle\phi^2\rangle_\text{ren}$, in free field theory. We numerically show that the identity perfectly holds in this case. In Fig.\ref{fig:ward} the solid line corresponds to $\partial_t \mathcal{E}_\text{ren}$ which is directly computed from the renormalized version of the energy density
\be\label{eq:Edensity}
\hspace{-4mm}
\mathcal{E}=\sigma_d\int\,dk\,k^{d-2}\left(\left|\partial_t u_k\right|^2+\left|\partial^z_i u_k\right|^2+m^{2z}(t)\left|u_k\right|^2\right),
\ee
and the data points correspond to $\frac{1}{2}\partial_t m^{2z}(t)\langle\phi^2\rangle_\text{ren}$ which are read from from Eq.\eqref{eq:FTmass} and Eq.\eqref{eq:phi2ren}.    

Before concluding we would like to emphasis that although the numerical results reported here correspond to $z=2$, we have checked all of our results for higher values of dynamical exponents in various dimensions.

\section{Conclusions}
We have shown that when the quench rate is fast compared to other scales in the theory, but still slow compared to the UV cut-off, the response of the system scales universally. The analysis even does not depend on the initial state of the system. The latter has a geometrical realization in the holographic side; the propagation of the response is restricted to a near-boundary region \footnote{In our holographic analysis we have considered asymptotic Lifshitz solutions of Einstein theory (in presence of matter fields) which makes the `causal' propagation of the response function inside a widen light-ray sensible. On the field theory side the same notion has been found via a $z$-dependent maximum group velocity for propagating modes in free theory \cite{MohammadiMozaffar:2018vmk}.}.

We have found that this universal bahavior is the case on one hand for strongly coupled theories and on the other hand for free theories. Although there is no proof, based on pertubative arguments in support of the same feature in interacting theories \cite{MMinProgress}, we believe that this scaling behavior is the case for generic interacting theories admitting a Lifshitz-like fixed point.

In relativistic case universal scaling is shown to be true not in the fast regime but for any rate \cite{Das:2016lla}. Extending our field theory analysis we also find that this is the case for Lifshitz theories not only for mass operators but also for correlators and for more complicated objects such as entanglement entropy. We will report these results in early future in \cite{MMinProgress} \footnote{Recently entanglement entropy for smooth quenches has been studied in these models \cite{EETE}.}.   

We have also found perfect agreement between our studies in the continuum and a latticized version of Lifshitz free theories in low dimensions \cite{MohammadiMozaffar:2017nri}.

\section*{Acknowledgements}
We would like to thank Diptarka Das and Tadashi Takayanagi for fruitful discussions and Sumit Das for correspondence. We are grateful to Rob Myers for carefully reading the draft and for his valuable comments. AM would like to thank the organisers of ``Quantum Information and String Theory 2019" workshop at YITP, Kyoto, where these results where presented. AM is generously supported by Alexander von Humboldt foundation via a postdoctoral fellowship.

\end{document}